\documentstyle[prl,aps,multicol,epsf]{revtex}
\newcommand{\be}{\begin{equation}}
\newcommand{\ee}{\end{equation}}
\newcommand{\bea}{\begin{eqnarray}}
\newcommand{\eea}{\end{eqnarray}}

\begin{document}
\draft 
\twocolumn[\hsize\textwidth%
\columnwidth\hsize\csname@twocolumnfalse\endcsname
\title{
{Charge Density Wave--Assisted Tunneling Between Hall Edge States}}
\author{L. Moriconi}
\address{Instituto de F\'\i sica, Universidade Federal do Rio de Janeiro,\\
C.P. 68528, Rio de Janeiro, RJ -- 21945-970, Brasil}
\maketitle
\begin{abstract}
We study the intra-planar tunneling between quantum Hall samples separated by a quasi one-dimensional barrier, induced through the interaction of edge degrees of freedom with the charge density waves of a Hall crystal defined in a parallel layer. A field theory formulation is set up in terms of bosonic (2+1)-dimensional excitations coupled to (1+1)-dimensional fermions. Parity symmetry is broken at the quantum level by the confinement of soliton-antisoliton pairs near the tunneling region.
The usual Peierls argument allows to estimate the critical temperature $T_c$, so that for $T > T_c$ mass corrections due to longitudinal density fluctuations disappear from the edge spectrum. We compute the gap dependence upon the random global phase of the pinned charge density wave, as well as the effects of a voltage bias applied across the tunneling junction.
\end{abstract}
\vskip1pc] 
\narrowtext

{\it{Introduction}}.
The existence of charge density waves (cdws) in quantum Hall systems, initially conjectured to arise in connection with cyclotron resonances and later on with the destruction of the fractional effect in disordered samples \cite{girvin}, has by now convincing experimental and theoretical ground \cite{koulakov,moessner,rezayi,lilly}. However, it has been recently found \cite{murthy,tesanovic} that for certain regimes at filling fractions $\nu > 2$, the spectral properties of charge density fluctuations at low wavenumbers may substantially differ from the usual theoretical expectations, formulated even before the discovery of the integer quantum Hall effect \cite{fukuyama}. Gapless longitudinal and transverse modes for these exotic ``Hall crystals" \cite{tesanovic} are predicted to be characterized by linear dispersion curves, likely to be rendered massive by the presence of disorder.

Our aim in this paper is to propose and study a double layer system conceived to probe the dynamics of cdws with two-dimensional lattice structure in the quantum Hall effect, under the assumption they are governed by a linear dispersion in the gapless limit. The proposed experimental construction is schematically depicted in fig.1. Two close quantum Hall layers
are coupled through the Coulomb interaction. A quantum Hall cdw state is defined in the upper sample (referred to as system I). The lower sample (system II) is an assemblage of two coplanar systems, separated by a tunneling junction of width $\sim \ell \sim 10^2$ \AA  {\hbox{ }}(the magnetic length in usual quantum Hall samples), which is surrounded by two sets of oppositely oriented chiral edge states (``left" and ``right" movers) \cite{mori}. The essential situation we will focus is the tunneling resonance that occurs when the Fermi level in system II is tuned so that the left and right chiral edge states become strongly coupled via the periodic structure of the cdw state in system I. Such a tunneling resonance is actually attainable due to the fortunate fact that the lattice parameter in the Hall cdw states is also close to $\ell$ \cite{koulakov,rezayi,murthy}.

Before embarking in the main considerations, a disclaimer is in order. It is reasonable to suppose that the experimental search for a Hall crystal should be done in principle with the help of standard approaches, like radio-frequency absortion and transport measurements, performed on single layer structures \cite{proc}. However, even if a clear experimental signature of the Hall crystal is found in that way, the investigation of its low-energy spectrum turns out to be a more involved problem. It should be clear that we do not intend to establish here {\it{the way}} to {{\it{detect}} a Hall crystal, but just advance {\it{a way}} to {\it{probe}} its properties.
\begin{figure}
\centerline{\epsfxsize = 6.0cm \epsfbox{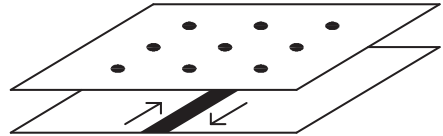}} \narrowtext\hspace{1.0cm}
\caption{The double layer system designed to probe low energy modes of a quantum Hall cdw state in the upper layer (system I) through the induced intra-planar tunneling which takes place between edge states in the lower layer (system II).}
\label{fig:1}
\end{figure}

The basic physical conditions necessary for the experiment are completely realistic. Quantum Hall layers separated by distances near $10^2$ \AA {\hbox{ }}are the subject of actual experimental and theoretical investigations (see the review \cite{girvin2}), whereas tiny (at the magnetic length scale) and precisely constructed quasi one-dimensional barriers were recently used by Kang {\it{et al}}. to study the tunneling between coplanar quantum Hall samples \cite{kang}. 

To address (and by no means exhaust) a discussion on the feasibility of the device sketched in fig.1, we find useful to comment on a few experimental aspects, having in mind modern nanostructure technology. Consider initially a Hall crystal which is hypothetically prepared in a very clean (high mobility) quantum well Al$_x$Ga$_{1-x}$As/GaAs/Al$_x$Ga$_{1-x}$As. The Hall crystal is believed to exist for some range of filling factors $2< \nu = nh/eB < 3$ at moderate Coulomb mixing between Landau Levels \cite{murthy}. Such a quantum well can be perfectly recreated in a double layer system, where it plays the role of system I. We may take the other quantum well (system II) to be the inversion layer at an adjoining Al$_x$Ga$_{1-x}$As/GaAs interface. The layers should be constructed within the cleaved edge overgrowth technique \cite{pfeiffer}, with the potential barrier in system II corresponding to a thin, atomically precise, region of an alloy of Al$_{x'}$Ga$_{1-x'}$As/AlAs inserted in the GaAs side before the cleave is made \cite{kang}. Usual voltage gates are placed above and below systems I and II, in order to control their electron densities $n_1$ and $n_2$ in an independent way \cite{dlayer}. It is clear that the determination of optimal values for quantum well and potential barrier thicknesses, the concentration of Si doping, Al fractions $x$ and $x'$, etc., is a matter of detailed quantum well engineering.

An advantage here over typical double layer tunneling experiments is that there is no need to attach separate ohmic contacts to the layers, a task generally accomplished through extra electrostatic gating \cite{pfeiffer2}. The layers can be connected in parallel, since $\sigma_{xx} =0$ in the Hall crystal state (of course, any external, in-plane electric field has to be small enough not to depin the cdw). In case there is some residual longitudinal conductivity contribution from the top layer, differential conductance peaks would reveal the presence of tunneling resonances in the bottom layer as well. The experimental pursue of the tunneling resonance can then be carried out in two essentially equivalent ways. Either by variating $n_2$ with all the other parameters fixed or, alternatively, by variating the magnetic field at fixed ratios $n_1/n_2$, to sweep the filling factors $2 < \nu_1 <3$ and $2n_2/n_1< \nu_2 < 3n_2/n_1$. Tunneling is expected to take place between edge states which belong to the lowest Landau level, even if $\nu_2 \neq 1$ \cite{kang}.

{\it{Definition of the tunneling model}}.
Let us imagine system II as a stripe in the $(x_1, x_2)$ plane, with $|x_1| < L/2$, where the limit $L \rightarrow \infty$ is eventually performed. The transverse magnetic field is $\vec B = -B \hat x_3$, and the tunneling interface is given by a triangular potential barrier $V(x_2)= -V_0 [2|x_2|/\delta - 1]$, defined for $|x_2| < \delta/2$. In our computations
the lateral size of the barrier, $\delta$, is assumed to be of the order of the magnetic length. A concrete range of values is $1$ meV $< eV_0 < 350$ meV for the potential barrier, and $1.5$ meV $< \hbar \omega_c < 15$ meV for the cyclotron energy.

The second quantized hamiltonian, which takes into account the interaction between system II and the cdw state in system I, is written in the Landau gauge as
$H=H_{\Phi}+H_{\Phi \xi}+H_\xi$, where
\bea
&H_\Phi& = \int d^2 \vec x \Phi^+ \{ {1 \over {2m}} [ (p_1 -eBx_2)^2 + p_2^2 ] + eV(x_2) \} \Phi  \ , \ \nonumber \\
&H_{\Phi \xi}& = \int d^2 \vec x d^2 \vec x' 
\rho(\vec x;\vec \xi) {e^2  \over {\epsilon [(\vec x - \vec x')^2 + d^2]^{1 \over 2}}}
\Phi^+(\vec x') \Phi (\vec x')
 \ , \ \nonumber \\
&H_\xi& = {u \over 2} \int d^2 \vec x [(\partial_t \vec \xi )^2 + c^2 (\partial_i \vec \xi)^2 + \omega_0^2 \vec \xi^2] \ . \
\label{ha}
\eea
Above, $\rho(\vec x;\vec \xi)$ is the deviation from the mean electron density in the Hall crystal. Density fluctuations around the periodic lattice are described by the distortion field $\vec \xi(\vec x,t)$. The non-local Coulomb coupling between electrons at the upper and lower samples is modeled through $H_{\Phi \xi}$, where $d$ in (\ref{ha}) is the distance between layers.

The harmonic hamiltonian $H_\xi$, which governs long wavelength fluctuations of the distortion field, is not written on a first-principles basis, otherwise it would necessarily introduce the coupling of the elementary charge motion to the external magnetic field. Rather, $H_\xi$ has to be regarded, in a ``Ginzburg-Landau" fashion, as the simplest phenomenological choice devised to satisfy a few basic physical properties: it must be i) a local and isotropic functional of $\vec \xi(\vec x,t)$, and ii) characterized by the fact that all modes are linearly dispersing in the gapless limit. The role played by the combination of magnetic field, Coulomb interaction and disorder is assumed to be implicit in the definition of coupling constants $u,c$ and $\omega_0$. In that sense, the gapless limit, $\omega_0 \rightarrow 0$, in $H_\xi$ holds for absence of disorder in system I. Furthermore, one may estimate the energy density parameter $uc^2$ from the Coulomb interaction within the magnetic length scale, that is, $uc^2 \sim e^2/\epsilon \ell^3  \equiv
r_s \hbar \omega_c / \ell^2$, where $r_s \geq 6$, following Murthy's numerical explorations \cite{murthy}.

There is some support to conjecture that $\omega_0 \rightarrow 0$ should indeed be taken for cdws realized in high mobility samples. From a study of the static competition between elastic, Coulomb and disordering interactions due to impurities, Fukuyama and Lee \cite{fukuyama} established a criterion to predict if there is or not a gap at wavenumber $q \rightarrow 0$ in the spectrum of general two-dimensional cdws. The Fukuyama-Lee criterion is rephrased as follows: let $n$ and $n_i$ be the electron and impurity densities, respectively, while the cdw has amplitude $\rho_0$ and lattice parameter $\sim 1 / \bar k$. An individual impurity is mimicked by the delta-potential $V(\vec r) = V_0 \delta^2 (\vec r)$, with $V_0 \sim e d_i$, where $d_i$ is of the order of the screening length. The two-dimensional cdw will have a diverging coherence length (positional order in practice) and gapless excitations if
\be
{ { V_0 \rho_0 \bar k^2  \sqrt{n_i}} \over {\pi e n^2 }} < 1 \ . \ \label{f-lee}
\ee
Considering that $\rho_0 < n \sim 1 / \ell^2$, $\bar k \sim 1/ \ell$ and $n_i \sim 1 / d_i \ell_p$, where $\ell_p$ is the mean free path (evaluated at zero magnetic field), we securely get (\ref{f-lee}) if $d_i / \ell_p \ll 1$.
Since $d_i$ is a few angstroms large and $\ell_p \simeq 10 \mu$m for a sample with mobility $\mu = 10^6$ cm$^2$/Vs, we expect (\ref{f-lee}) to be satisfied.

With no loss of important details, the cdw state $\rho(\vec x;0)$ will be taken to be a square lattice with spatial period $\pi / \sqrt{2} \bar k$ . We may write
\bea
\rho (\vec x ; \vec \xi) &=& {1 \over 2} \rho_0 J \{ \cos[2 \sqrt{2} \bar k \hat n_1 \cdot ( \vec x+ \vec \xi 
+ \hat x_2 \theta/2 \bar k)] \nonumber \\
&+& \cos[ 2 \sqrt{2} \bar k \hat n_2 \cdot (\vec x+ \vec \xi+ \hat x_2 \theta /2 \bar k)] \} \ , \ \label{cdw}
\eea
where $J = \det [ \delta_{\alpha \beta} + \partial_\alpha \xi_\beta ]$ is the jacobian which guarantees charge conservation, that is, $\int d^2 \vec x \rho (\vec x; \vec \xi) = 0$. Crystal axes are given by $\hat n_1$ and $\hat n_2$. We assume small and smooth distortions (at the magnetic length scale) of the square lattice. The global phase $\theta$ in (\ref{cdw}) gives finite translations of the pinned cdw along the $x_2$ direction. It is not necessary to consider a phase parameter for the perpendicular direction, since the total hamiltonian is invariant under shifts of $x_1$. The phase $\theta$ is expected to take random values each time the two-dimensional cdw is crystallized, in view of the extreme sensitivity of the pinning process to small perturbations.

The careful reader may have already noticed that we neglected, in (\ref{ha}), the Coulomb interaction between the cdw and the charge depleted region at the barrier in the bottom layer. Actually, due to translation symmetry along the $x_1$ axis, the charge depleted region generates a potential $\tilde V(x_2)$ in the top layer. Therefore, we should have added, in principle, the correction term
\be
\Delta H = e \int d^2 \vec x \tilde V(x_2) \rho (\vec x ; \vec \xi) \label{strip}
\ee
to $H_{\phi \xi}$. However, taking now an homogenous distortion $\vec \xi$ and using (\ref{cdw}) and (\ref{strip}) with periodic boundary conditions, the integration over $x_1$ implies, for general cdw orientations, that $\Delta H \rightarrow 0$ in the low wavenumber limit (that is, $\Delta H$ is a boundary term). It is solely in the special case when one of the crystal axes is parallel to the $x_1$ direction, as in fig.1, that we may have $\Delta H = \Delta H [ \xi_2] \neq 0$, so that the charge depleted region will favour some displacement of the cdw along the $x_2$ direction. 

For the benefit of a direct exposition, instead of developing computations for arbitrary axes directions $\hat n_1$ and $\hat n_2$, we will consider fully isotropic cdw fluctuations, with $\hat n_1 = (\hat x_1 + \hat x_2)/\sqrt{2}$ and $\hat n_2 = (\hat x_1 - \hat x_2)/ \sqrt{2}$, i.e.,
\be
\rho (\vec x ; \vec \xi) = \rho_0 J \cos[2\bar k ( x_1+ \xi_1)] \cos[ 2\bar k (x_2+\xi_2)+\theta] \ . \ \label{cdw2}
\ee
When appropriate, we comment on the modifications introduced by different choices of the crystal axes.

In order to define an expansion for the second quantized fermion operator, we work in the lowest Landau level approximation, taking for the Hilbert space basis the one-particle eigenstates of $H_\Phi$ with $V_0 = 0$. Up to a normalization constant,
\be
\Phi (x_1, x_2) 
= \sum_{n} a_n \exp[ik_nx_1 - {1 \over {2 \ell^2}}(x_2 - k_n \ell^2)^2 ] \ . \ \label{secq}
\ee
Imposing the periodic boundary condition $\Phi(x_1,x_2)= \Phi(x_1+L,x_2)$, one gets the discrete set of wavenumbers $k_n = 2 \pi n/ L$, with $n$ integer. The operator $a_n^+$ creates a particle which is free to move along the $x_1$ direction, but is localized around $\langle x_2 \rangle = k_n \ell^2$. It is clear from the form of (\ref{cdw2}), that a tunneling resonance occurs at the Fermi level which crosses the energy band generated by the potential barrier at $k_n = \pm \bar k$. Our task, thus, is to obtain a theory for the edge degrees of freedom with quantum numbers around these special values of $k_n$. The starting point is to rename operators as
\be
a_n = ({ {2 \pi} \over L})^{{1 \over 2}}
\left \{
\begin{array}{cl}
a_R (k + \bar k) & \mbox{, if $n<0$} \\
\\
a_L (k - \bar k) & \mbox{, if $n>0$} 
\end{array}
\right. \ . \ \label{chiral}
\ee
The indices $R$ and $L$ denote chiral components defined in a reference frame co-moving respectively with the $- \bar k$ and $\bar k$ states (when the barrier electric field is ``turned on").
Substituting (\ref{cdw2})-(\ref{chiral}) in $H_\Phi$ and $H_{\Phi \xi}$, we perform the integrations over $x_2$, taking $\vec \xi (x_1,x_2) \simeq \vec \xi (x_1,0)$ close to the tunneling region. It is also convenient to introduce the Dirac spinor
\be
\psi = {1 \over \sqrt{2 \pi}} \int dk \exp(ikx)
\left[
\matrix{ a_R (k) \cr
         a_L (k) } \right] \ , \ \label{dirac}
\ee
and the chiral representation of gamma matrices, written in terms of Pauli matrices as $\gamma^0 = \sigma_1$, $\gamma^1 = -i \sigma_2$ and $\gamma^5 = \sigma_3$. We find then
\bea
H &&= H_\xi+ \int dx  [ - iv \bar \psi \gamma^1 \partial_1 \psi \nonumber \\
&&+{g \over 2} \cos(2 \bar k \xi_2 + \theta) \bar \psi \exp(2i\bar k \xi_1 \gamma^5) \psi ] \ , \ \label{ha2}
\eea
where $v=e V_0 \ell^2/\delta$ is the drift velocity near the barrier, and $g = e^2 \pi \rho_0 \exp( - \bar k d)
/ \epsilon \bar k$
parametrizes the Coulomb coupling between layers. Note that $H_{\Phi \xi}$ is mapped into local terms in (\ref{ha2}), an approximation related to the smoothness properties of the distortion field. 

The above hamiltonian describes the interaction of (1+1)-dimensional Dirac fermions -- the edge excitations -- with (2+1)-dimensional vector fields -- the cdws -- along the line $x_2 =0$. Two-body interaction effects within system II will not be considered, which may be addressed in a luttinger liquid framework, once they are likely to just renormalize the bare (input) parameters of the fermion theory, as the gap (its size and position) and the drift velocity \cite{mitra,lee,sachdev}.

As is well-known, disorder is expected to broaden and reduce any tunneling resonance peak. It is not difficult to show disorder may be effectively incorporated in (\ref{ha2}) as a random gauge field coupled to the edge states, in a way suitable of analysis via the replica formalism \cite{bocquet}. While leaving a deeper study of disorder for further work, we just point, from previous experiments \cite{kang} that disorder broadenening, if any, is typically of the order of $1$ meV, which sets an energy resolution for the observation of many of the results derived here. 

Regarding alternative cdw orientations, it is worth noting there will be in general {\it{two}} resonant Fermi levels, which correspond to the two periodic functions in (\ref{cdw}). The linearization procedure around the resonances leads now to coupling terms like
\be
\sum_{p=1}^2 g_p \bar \psi \exp[i 2 \sqrt{2} \bar k \hat n_p \cdot (\vec \xi + \hat x_2 \theta / 2 \bar k)\gamma^5] \psi 
\ . \
\label{coupl}
\ee
The ``degenerate case", (\ref{ha2}), follows from $g_1=g_2=g/4$ with $\hat n_1 = (\hat x_1 + \hat x_2)/\sqrt{2}$ and $\hat n_2 = (\hat x_1 - \hat x_2)/ \sqrt{2}$. As $\hat n_1$ and $\hat n_2$ are rotated, one of the coupling constants $g_1$ or $g_2$ gets comparatively smaller. This is in fact the behaviour related to a more realistic cdw profile, which is peaked in two-dimensional Fourier space at $2 \sqrt{2} \bar k \hat n_1$ and $2 \sqrt{2} \bar k \hat n_2$, but departs from the idealized harmonic form (\ref{cdw}). If the rotation angle is large enough, we may discard one of the $g's$ in (\ref{coupl}), implying physical quantities become independent of $\theta$. That is what happens when $\hat n_1$ is parallel to the tunneling barrier. Notwithstanding the anisotropic cdw fluctuations which also appear in this situation, tunneling will be due to the interaction of edge states with $\xi_1$, whereas any extra-term $\Delta H$ introduced to accommodate the effect of the charge depleted region will have to do with transverse fluctuations of the distortion field. As a consequence, the coupling between the top layer and the charge depleted region does not play a relevant role in the tunneling process.

The rationale in focusing the degenerate case in detail relies on its rich vacuum structure, qualitatively similar (or even more general) than the ones found for diverse cdw orientations.

{\it{Vacuum structure of the non-local effective fermion model}}.
Our main interest is to determine the fermion mass -- the tunneling gap -- which could be experimentally determined from the Fermi level positions of zero-bias differential conductance peaks, for instance.
A naive inspection of the hamiltonian (\ref{ha2}}), ``freezing" the bosonic degrees of freedom in some homogenous configuration $\vec \xi$, suggests that the fermion mass would be $m = g |\cos( 2 \bar k \xi_2+ \theta)|/2$, since the constant $\xi_1$ can be eliminated by a chiral rotation of the fermion fields. However, this simple argument is completely misleading, since it neglects fluctuations of the distortion field, which are particularly relevant in planar systems.

The central point of our analysis, to be performed in the framework of finite temperature field theory \cite{das}, is the elimination of $\vec \xi$ from the partition function by means of a standard projection procedure. The computational strategy is based on the cumulant expansion for the perturbative hamiltonian piece of $O(g)$ in (\ref{ha2}), in terms of averages over fluctuations of $\vec \xi$. We will be able to derive in this way an effective (non-local) (1+1)-dimensional fermion model, which describes the interaction between edge states.

The partition function is written, in a self-evident notation, as
\be
Z = \int D \vec \xi D \bar \psi D \psi \exp [ - S_{\xi}- S_{\psi} - S_{\xi \psi}] \ . \ \label{partf}
\ee
We get, from the second-order cumulant expansion,
\be
Z = Z_0 \int D \bar \psi D \psi \exp[ -S_\psi - \Delta S_\psi] \ , \ \label{partf2}
\ee 
where
\be 
\Delta S_\psi = \langle S_{\xi \psi} \rangle_0 + {1 \over 2} [ \langle S_{\xi \psi}^2 \rangle_0 - \langle S_{\xi \psi} \rangle_0^2 ] \ , \ \label{ds-psi}
\ee
with $\langle (...) \rangle_0$ standing for the statistical average taken with the partition function $Z_0 = \int D \vec \xi \exp[-S_{\xi}]$. We have
\be
S_\xi = {u \over 2} \int d \tau d^2 \vec x [(\partial_\tau \vec \xi )^2 + c^2 (\partial_i \vec \xi)^2 + \omega_0^2 \vec \xi^2] \label{s-xi}
\ee
and
\be
S_{\xi \psi} = {g \over 2} \int d \tau dx \cos(2 \bar k \xi_2 + \theta) \bar \psi \exp(2i\bar k \xi_1 \gamma^5) \psi \ . \ \label{s-xipsi}
\ee
The integration over imaginary time is restricted to the interval $0 < \tau < \beta \equiv 1 /T$, and periodic and antiperiodic boundary conditions are defined, as usual, for the boson and fermion fields, respectively. The translation invariant bosonic two-point correlation function, evaluated on the line $x_2=0$, is $\langle \xi_i(0,0,0) \xi_j(x,0,\tau) \rangle_0 = \delta_{ij} G(x,\tau)$, with
\bea
&&G(x, \tau)= \nonumber \\
&=&{1 \over {2\pi \beta}} 
\sum_{n} \int dk
{{(2 u c^2)^{-1}\exp[i( k x +\omega_n \tau)]} 
\over {[k^2 + (\omega^2_n +\omega_0^2)/ c^2 ]^{1 \over 2}}}  \ , \ \label{propg}
\eea
where $\omega_n = 2 \pi n / \beta$ are the Matsubara frequencies. To obtain the statistical averages in (\ref{ds-psi}), it is only necessary to know that
\bea 
&& \langle \sin[2 \bar k \xi_i] \rangle_0 =  \langle \sin[2 \bar k \xi_i] \cos[2 \bar k \xi_j] \rangle_0 = 0 \ , \ 
\nonumber \\
&&  \langle \cos[2 \bar k \xi_i] \rangle_0 = \cosh[ 2 \bar k^2 G(0)] \ , \ \nonumber \\
&&  \langle \cos[2 \bar k \xi_i(0,0,0)] \cos[2 \bar k \xi_j(x,0,\tau)] \rangle_0 = \delta_{ij} \exp[-4 \bar k^2
G(0)] \nonumber \\
&&\times \cosh[ 4 \bar k^2 G(x,\tau)]+(1 - \delta_{ij})\cosh^2[ 2 \bar k^2 G(0)] \ , \ \nonumber \\
&&  \langle \sin[2 \bar k \xi_i(0,0,0)] \sin[2 \bar k \xi_j(x,0,\tau)] \rangle_0 = \delta_{ij} \exp[-4 \bar k^2 G(0)] 
\nonumber \\
&&\times \sinh[ 4 \bar k^2 G(x,\tau)]  \ . \  \label{c-funct}
\eea
The exact results in (\ref{c-funct}) can be approximated by more convenient expressions, if we note that all of them involve
functions of $\bar k^2 G(x,\tau) = O(\xi^2/ \ell^2) \ll 1$. We may take, in practice, the leading order contributions:
\bea
&&  \langle \cos[2 \bar k \xi_i] \rangle_0 \simeq \langle \cos[2 \bar k \xi_i(0,0,0)] \cos[2 \bar k \xi_j(x,0,\tau)] \rangle_0 \simeq 1 \ , \ \nonumber \\
&&  \langle \sin[2 \bar k \xi_i(0,0,0)] \sin[2 \bar k \xi_j(x,0,\tau)] \rangle_0 \simeq 4 \bar k^2 G(x,\tau)  \ . \  \label{c-funct2}
\eea
A direct computation gives, then, 
\bea
\Delta S_\psi =  && {i g \over 2} \cos \theta \int d \tau dx \bar \psi \gamma^5 \psi + {{ g^2 \bar k^2} \over 2} 
\int d \tau d \tau' dx dx'  \nonumber \\
&& \times G(x-x',\tau - \tau') [ \cos^2 \theta (\bar \psi \psi)_{x \tau} (\bar \psi \psi)_{x' \tau'} 
\nonumber \\
&&-\sin^2 \theta (\bar \psi \gamma^5 \psi)_{x \tau} (\bar \psi \gamma^5 \psi)_{x' \tau'} ] \ . \ \label{ds-psi2}
\eea
Therefore, we have to study, from now on, the (1+1)-dimensional fermion theory, with partition function (\ref{partf2}) and non-local couplings given in (\ref{ds-psi2}).
The most natural approach, following an analogy with the problem of BCS superconductivity, is to devise a Hartree-Fock computation of the gap equation. In the present context, we define the non-local order parameters
\bea
\varphi_1(x,\tau) &&= \int d \tau' dx' G(x-x',\tau - \tau') (\bar \psi \psi)_{x' \tau'} \ , \ \nonumber \\
\varphi_2(x,\tau) &&= \int d \tau' dx' G(x-x',\tau - \tau') (\bar \psi \gamma^5 \psi)_{x' \tau'} \ . \ \label{orderp}
\eea 
If the vacuum expectation value of $\vec \varphi \equiv \varphi_1 \hat x_1 + \varphi_2 \hat x_2$ is non-vanishing, we find, from (\ref{ds-psi2}) and (\ref{orderp}), the fermion mass
\be
m=g[\bar k ^2 \cos^2 \theta \langle \varphi_1 \rangle^2 + 
({1 \over 2} \cos \theta  -  \bar k \sin \theta \langle \varphi_2 \rangle)^2]^{1 \over 2} \ . \ \label{masseq}
\ee

A fundamental aspect of the effective fermion theory obtained after the integration over the distortion fields regards its parity symmetry properties. At the tree-level there is invariance under the discrete mapping $x_1 \rightarrow -x_1$ and $\psi \rightarrow \gamma^0 \psi$, which implies that $\varphi_1 \rightarrow -\varphi_1$. One could then state that $\langle \varphi_1 \rangle = 0$ is necessarily verified, since discrete symmetries cannot be broken in one spatial dimension. However, such a general theorem strictly holds for local hamiltonians. As we discuss below, the non-local coupling between fermions renders in fact parity symmetry breaking possible when quantum corrections are taken into account.

We may investigate the vacuum phases of the fermion theory by expressing the partition function as a functional integration over configurations of $\vec \varphi$. The basic difficult in doing so is related to the evaluation of the fermion determinant, which, however, admits a loop-expansion. At the one-loop level \cite{coleman} we get the partition function
\be
Z = \int D \vec \varphi \exp[-S_\varphi] \ , \ \label{partfunc3}
\ee
where
\bea
S_\varphi &&=  \int d \tau dx \{ uc^2 \sum_{i=1}^2 \varphi_i [\partial_\tau^2 +\partial^2 + \omega_0^2 / c^2]^{1 /2}  \varphi_i  \nonumber \\
&&  +( \sigma^2 / 4 \pi v ) [ \ln  ( \sigma^2 /  \Lambda^2) -1 ] \} \ . \ \label{action-phi}
\eea
Above, $\sigma^2 = g^2[\bar  k^2 \cos^2 \theta \varphi_1^2 + ({1 \over 2} \cos \theta - \bar k \sin \theta \varphi_2)^2]$ and $\Lambda$ is the ultraviolet cutoff, which may be regarded to be of the order of $e V_0$. 
The effective potential is, thus,
\be
V_{\hbox{eff}}(\varphi_1,\varphi_2) =
uc\omega_0 \vec \varphi^2 + {{ \sigma^2} \over   {4 \pi v}} \left [ \ln  { \sigma^2 \over  \Lambda^2} -1 \right ]
\ . \ \label{effecpot}
\ee
It follows from (\ref{effecpot}) that there are two types of ground states, which we refer to as ``type-A" and ``type-B". Type-A ground state breaks parity symmetry and is associated to the fermion mass
\be
m= \Lambda \exp(-  {{\bar c \omega_0} \over { \bar k^2 g^2 \cos^2 \theta}} ) \  , \ \label{mass}
\ee
where $\bar c = 2 \pi v u c$. Such a vacuum is realized for a specific pinning parameter $\theta$ whenever $m > g| \cos \theta / 2 (\tan^2 \theta -1)|$. Vacuum expectation values are
\bea
&&\langle \varphi_1 \rangle =  \pm [{m^2 \over {\bar k^2 g^2 }} \sec^2 \theta - 
\langle \varphi_2 \rangle^2 \cot^2 \theta]^{1 \over 2} \ , \ \nonumber \\
&&\langle \varphi_2 \rangle= {{\tan \theta} \over { 2 \bar k (\tan^2 \theta -1)}} 
\ . \ \label{vev}
\eea

The parity symmetric type-B ground state is given on its turn by $\langle \varphi_1 \rangle=0$ with $\langle \varphi_2 \rangle$ being obtained from 
\be
{{2 \bar c \omega_0} \over {\bar k g \sin \theta}} \langle \varphi_2 \rangle = \chi \ln  {\chi^2 \over \Lambda^2} \ , \ \label{transc}
\ee
where $\chi = g (\cos \theta -2 \bar k \langle \varphi_2 \rangle \sin \theta)/2$. The fermion gap is $m= | \chi |$. With the exception of $\theta = \pi/2$ and $3 \pi/2$, there is no vacuum degeneracy here. Type-B ground state is always the correct choice for some range of pinning phase parameters, and in some situations, as in the weak or strong coupling limits of $g$, it holds in fact for any value of $\theta$.

To study the quantum stability of type-A vacuum, we employ a variational strategy, where dilute soliton-antisoliton configurations interpolating between the degenerate vacuum states $\langle \varphi_1 \rangle \equiv \pm \varphi_0$, are showed not to disorder the system at low temperatures. A variational soliton of size $\sim a$ is constructed in a simple way through the convolution of the step and gaussian functions. Just define
\bea
\varphi^s_1(x) &=& {{2 \varphi_0} \over {a \sqrt{\pi}}} \int d x'  { {x'} \over {|x'|}}  \exp[ -{{4 (x-x')^2} \over a^2}] 
\nonumber \\
&=& \varphi_0 \int {{dk} \over {\pi i}}
{{k \exp(ikx-k^2 a^2)} \over {k^2 +\eta^2}} \ . \ \label{sol}
\eea
where $\eta \rightarrow 0$. The antisoliton is, of course, $\varphi^{\bar s}_1(x) = - \varphi^s_1(x)$.
The soliton energy, subtracted by the ground state level, is
\bea
\Delta E_s &=& {{2 u c^2 \varphi_0^2} \over \pi} \int {{dk} \over k^2}
\left [ \left (k^2 + {\omega_0^2 \over c^2}\right )^{1 \over 2} - {\omega_0 \over c} \right ]
\exp(-2 k^2 a^2) \nonumber \\
&+& \int dx[ V_{\hbox{eff}}(\varphi^s_1,\langle \varphi_2 \rangle) - 
V_{\hbox{eff}}(\varphi_0,\langle \varphi_2 \rangle)] \ . \ \label{s-energy}
\eea
While the second term on the rhs of (\ref{s-energy}) may be approximated by a linear expression, $\gamma a$, with $\gamma > 0$, the first term diverges for $a \rightarrow 0$ in the ultraviolet limit and vanishes for $a \rightarrow \infty$, implying there is necessarily a minimum of $\Delta E_s$ at some $a=\bar a$. Thus, the existence of a soliton state is verified. 

Consider now a soliton-antisoliton pair separated by a distance $x_0 \gg \bar a$, which may be written as the superposition $\varphi^{s \bar s}_1 (x) = \varphi^s_1(x) + \varphi^{\bar s}_1(x-x_0) - \varphi_0$. Its energy is
\bea
&\Delta& E_{s \bar s} = 2 \gamma \bar a + {{8 u c^2 \varphi_0^2} \over \pi}
\int {{dk} \over k^2} \nonumber \\
&&\times \sin^2 {{k x_0} \over 2} \left [ \left (k^2 + {\omega_0^2 \over c^2} \right )^{1 \over 2} - {\omega_0 \over c} \right ] \exp(-2 k^2 \bar a^2)  \ . \ \label{sas-energy}
\eea
Taking $\omega_0$ finite, it follows that $\Delta E_{s \bar s}$ approaches a constant in the large $x_0$ limit. In this case, entropy effects dominate the free energy and type-A vacuum is disordered, that is, $\langle \varphi_1 \rangle =0$. However, if $\omega_0 \rightarrow 0$ at fixed $x_0$, one gets the asymptotic result, $\Delta E_{s \bar s} \sim \ln(x_0 / \bar a)$. Energy and entropy fluctuations have similar logarithmic dependences, and a direct application of the Peierls argument \cite{peierls} gives the critical temperature $T_c ( \theta ) \sim u c^2 \varphi_0^2$. With $r_s = 6$, $\hbar \omega_c = 1.5$ meV, and assuming $\varphi_0 \sim 0.05 \ell$, for instance, we find $T_c \sim 0.3$ K. In the disordered phase above $T_c$, similarly to the finite $\omega_0$ case, the longitudinal bosonic excitations along the interface do not contribute to the fermion mass renormalization (up to one-loop level). 

Observe that for $\omega_0 \rightarrow 0$, the soliton energy diverges in the infrared limit, but the soliton-antisoliton energy is finite. The soliton-antisoliton confinement below $T_c$ is analogous to the binding of vortices in the Kosterlitz-Thouless model \cite{kt}. As a matter of fact, the logarithmic interaction of kinks in a Ising-like model was studied long ago by Anderson, Yuval and Hamann as a way to model the Kondo effect \cite{anderson}. The renormalization group flows are essentially the same as the ones for the Kosterlitz-Thouless transition.

{\it{Finite voltage bias effects}}.
If a small voltage bias $V_b$ is applied across the quasi one-dimensional interface, the potential barrier becomes locally modified by the addition of a linear potential, viz.,
\be
V(x_2) = -V_0 [{{2|x_2|} \over {\delta}} - 1]- {{V_b x_2} \over {2 \delta}} \ . \ \label{pot}
\ee
As a consequence, the chiral fermion components will have different drift velocities (however, the average drift is kept constant). It is necessary to modify the hamiltonian (\ref{ha2}) according to
\be
\bar \psi \gamma^1 \partial_1 \psi \rightarrow 
\bar \psi \gamma^1 \partial_1 \psi - {V_b \over V_0} \bar \psi  \gamma^0 \partial_1 \psi \ . \ \label{drift}
\ee
Two distinct effects come into play. First, the modification of the fermion dispersion profile leads to a shift of the Fermi level where the tunneling resonance occurs. Let $E_f$ be the Fermi level for the tunneling resonance at zero bias. At a finite bias $V_b$, a purely geometrical reasoning shows the energy position of the resonance moves up:
\be
E_f'= E_f+(e V_0 - E_f) {V_b^2 \over V_0^2} \ . \ \label{shift}
\ee
Second, the one-loop computation may be performed once again, yielding up to $O(V_b^2)$ a renormalization of the cutoff $\Lambda$ which appears in the expression of the effective potential. Taking $\zeta \equiv \Lambda / \sigma_0$, where $\sigma_0$ denotes the gap evaluated at zero bias, we find $\Lambda$ is replaced by $\tilde \Lambda = \Lambda \exp(\kappa V_b^2/V_0^2)$, with
\be
\kappa = {2 \over \pi} (1 + \zeta^2 )^{1 \over 2}
\arctan [ (1 +\zeta^2)^{-{1 \over 2}} ]-{\zeta \over 2} 
\ . \ \label{d-const}
\ee
In that way, the gap for type-A vacuum is enhanced at finite voltage bias. The same phenomenon happens for type-B vacuum when $\omega_0 \rightarrow 0$. In that limit, $\sigma_0 \rightarrow \Lambda$ for both types of vacua, so that $\kappa \simeq 0.054$.

An interesting experiment involving finite bias effects in the clean regime $\omega_0 \rightarrow 0$ is as follows. Imagine the double layer system is initially set in type-B ground state (through $\Lambda < g$). The fermion gap is then $m \sim \Lambda$, irrespective of temperature and the pinning parameter $\theta$. As the voltage bias $V_b$ is increased, the effective cutoff $\Lambda$ gets enhanced as discussed above, so that at some point the quantum-disordered vacuum of type A arises, characterized by $\langle \varphi_1 \rangle =0$ at $\theta \simeq 0$ and $\pi$. Using (\ref{masseq}) and (\ref{vev}), we get, at such values of $\theta$, the gap $m \sim g/2$. Furthermore, since the critical temperature $T_c(0)=T_c(\pi) \propto \varphi_0^2 \sim (4 \Lambda^2/g^2 -1)$ grows with $V_b$, a low temperature regime may be eventually attained, where type-A vacuum becomes ordered, and the gap gets back to its original size $\Lambda$.

{\it{Conclusions}}.
We studied possible vacuum phases which appear in the coplanar tunneling between Hall edge states, as induced through the interaction with cdws in a double layer system. Quantum and thermodynamical critical transitions are found, yielding a rich phenomenological stage that could be explored to probe the dynamics of cdws in the quantum Hall effect. The above results are intimately related to the existence of gapless and linearly dispersing cdw modes. The usual gapless $q^{3/2}$ dispersion \cite{fukuyama}, for instance, would lead to weakly coupled soliton-antisoliton configurations and to permanently quantum-disordered vacuum states. Further interesting work could be concerned with alternative cdw orderings (stripes), and the tunneling between fractional Hall edge states. 
\smallskip

The author thanks D.G. Barci for calling his attention to ref. \cite{mitra}. This work has been partially supported by CNPq.

\end{document}